\documentclass[12pt]{article}
\usepackage{color}

\usepackage{epsfig,epstopdf}
\begin{document}
\newcommand{\bfi}{\it}
\newcommand{{\bfr}}{\mathbf{r}}
\newcommand{\beq}{\begin{equation} }
\newcommand{\eeq}{\eeq{equation} }
\newcommand{\comment}[1]{\vspace{1mm}\par
\framebox{\begin{minipage}[c]{.95 \textwidth} \rm #1
\end{minipage}}\vspace{1 mm}\par}
\newcommand{\rem}[1]{}
\newcommand{\remfigure}[1]{#1}
\newcommand{\neu}[2]{ \marginpar{\fbox{\color{red}\bf #1}}
                      {\color{blue}\bf #2}}
\newcommand{\rhobar}{\overline{\rho}}
\newtheorem{thm}{Theorem}

\title{
\vbox to 0pt {\vskip -1cm \rlap{\hbox to \textwidth {\rm{\small
FOR SUBMISSION TO {\it PHYSICA D.}
\hfil Today's version $\quad$  \today}}}}
Clumps and patches in self-aggregation of finite size particles}
\author{
Darryl D. Holm${}^{1,2}$ and Vakhtang Putkaradze${}^{3}$\\
\\${}^{1}$Theoretical Division\\
Los Alamos National Laboratory, Los Alamos, NM 87545, USA
\\{
email: dholm@lanl.gov}\\
and
\\${}^{2}$Mathematics Department\\
Imperial College London, SW7 2AZ, UK
\\{
email: d.holm@imperial.ac.uk}\\
\\${}^{3}$Department of Mathematics and Statistics\\
University of New Mexico,  Albuquerque, NM 87131-1141
\\{
email: putkarad@math.unm.edu}
}
\maketitle

\begin{abstract} \noindent
New model equations are derived for dynamics of self-aggregation of
finite-size particles. Differences from standard Debye-H\"uckel
\cite{DeHu1923} and Keller-Segel  \cite{KeSe1970} models are: a) the
mobility $\mu$ of particles depends on  the locally-averaged particle
density and b) linear diffusion acts on that locally-averaged particle
density. The cases both with and without diffusion are considered here.
Surprisingly, these simple modifications of standard models allow progress
in the analytical description of evolution as well as the complete analysis
of stationary states. When $\mu$ remains positive, the evolution of
collapsed states in our model reduces exactly to finite-dimensional
dynamics of interacting particle clumps. Simulations show these collapsed
(clumped) states emerging from smooth initial conditions, even in one
spatial dimension.  If $\mu$ vanishes for some averaged density,
the evolution leads to spontaneous formation of 
\emph{jammed patches} (weak solution with density having compact support).
Simulations  confirm that a combination of these patches forms the final
state for the system.  

\end{abstract}
\tableofcontents
\section{Historical perspective of  continuum models of
self-aggregation} Many fields of physics, chemistry and
biology deal with the problem of describing the evolution of the
macroscopic density of a large number of particles, which 
self-consistently  attract each other over distances large compared to
their mean separation. A related paradigm arises in biosciences,
particularly in chemotaxis: the study of the influence of chemical
substances in the environment on the motion of mobile species  which
secrete these substances. One of the most famous among such models is the
Keller-Segel elliptic-parabolic system of partial differential
equations \cite{KeSe1970}, which was introduced to explore the effects of
nonlinear cross diffusion in the formation of aggregates and
patterns by chemotaxis in the aggregation
of the slime mold {\it Dictyostelium discoidium}. The Keller-Segel model 
consists of two strongly coupled reaction-diffusion equations,
\rem{
\begin{eqnarray*}\label{Original-Keller-Segel}
u_t&=&{\rm div}(k_1(u,v)\nabla u - k_2(u,v)\nabla v)
\,,\\
\epsilon v_t&=&k_c\Delta v - k_3(v)v + f(v)u
\,.
\end{eqnarray*}
}
\begin{equation} \label{Keller-Segel}
\rho_t+ \mbox{div}  \mathbf{J} =0
\hspace{1cm}
\mathbf{J}
= \rho \mu(\Phi) \nabla \Phi - D\nabla \rho
\hspace{1cm}
\epsilon \Phi_t+L \Phi =\gamma \rho
\,,
\end{equation}
expressing the coupled evolution of concentration of
organisms (density) $\rho$ and concentration of chemotactic
agent (potential) $\Phi$. These
challenging nonlinear equations were posed for chemotaxis as an
initial value problem with
Neumann boundary conditions
initial data. The constants  $\gamma,D,\epsilon>0$  are assumed to be positive 
and $\mathbf{J}$ is the flux of organisms (particles). The linear
operator $L$ is taken to be positive and symmetric. For example,
one may choose it to be the Laplacian $L=-\Delta$ or the Helmholtz
operator, $L=1-\alpha^2 \Delta$. For the moment, the functional
dependence of the mobility $\mu$ will be left unspecified.%
\footnote{One recovers the precise expression of Keller and
Segel \cite{KeSe1970} by rewriting the mobility term as $\mu(\Phi)
\nabla \Phi=\nabla \chi(\Phi)$, where $\chi(\Phi)$ is the
``sensitivity function." The development below introduces a
different functional dependence in the mobility $\mu$.} The
``simplified" Keller-Segel model is obtained by setting 
$\epsilon=0$ in (1).  Because of its fascinating mathematical
properties -- such as finite-time
concentration of the density $\rho$ into Dirac delta-function
singularities, starting from smooth initial conditions --  and
its relevance to chemotaxis, the Keller-Segel system has attracted
sustained interest in mathematical biology.  To avoid
excessive citation here, we shall only refer to the review
article \cite{Ho2003} where over 150 references on the
role of (\ref{Keller-Segel}) in the chemotaxis problem may be
found.

The applications of equations in the form of system
(\ref{Keller-Segel}) in physics and chemistry may be traced back
into the 19th century. The history of these equations is a story
of recurrent rediscovery, each time as a leading order
description of singularity formation, occurring as the tendency
to diffuse is eventually overwhelmed by long-range forces of
attraction. Quite likely, their story has not finished yet and
these equations will be needed again for the insights they
provide into nonlinear multiscale dynamics in the competition
between coalescence and diffusion. 

Three main modeling steps figure in the derivation of (\ref{Keller-Segel})
and similar equations. First, the flux $\mathbf{J}$ must be modeled in
terms of the local particle density, its spatial gradient and the
gradient of the potential. Second, an equation must be formulated
for determining the potential $\Phi$ from the local density
$\rho$. Third,  conservation of mass, charge, or particle number
must be introduced as the evolution equation for density.

Historically, it seems that Debye and H\"uckel in 1923 were
the first to put all three of these modeling steps
together. They derived the evolutionary system
(\ref{Keller-Segel}) in their article \cite{DeHu1923} on the
theory of electrolytes. In particular, the simplified model with
$\epsilon=0$ in (\ref{Keller-Segel}) may be found as equations
(2) and (2$\,^\prime$) in Debye and H\"uckel
\cite{DeHu1923}.  Consequently, the simplified evolutionary
system (\ref{Keller-Segel}) with $\epsilon=0$ may also be
called  the Debye-H\"uckel equations.

The Debye-H\"uckel equations (\ref{Keller-Segel}) with
$\epsilon=0$ also appeared very early in astrophysics as the
Poisson-Smoluchowski system, a particular Fokker-Planck equation
describing a self-attracting reinforced random walk of
particles  (with mass, but without inertia) diffusing by
local collisions and drifting against friction under their mutual
long-range  attraction. As is well known, gravitational force
leads to collapse in this case, when mutual attraction finally
prevails over diffusion and friction. See Chandrasekhar (1939)
\cite{Ch1939} for an exhaustive historical review from
the viewpoint of stellar formation.

Interest in the Debye-H\"uckel system was recently
revived, when the ``Nernst-Planck'' (NP) equations
in the same form as (\ref{Keller-Segel}) re-emerged in the
biophysics community, for example, in the study of ion transport
in biological channels.  In this case, the flux
$\mathbf{J}$ is called the  Nernst-Planck (NP)
particle flux for the ion current density
$\mathbf{J}$ depending linearly on the
gradients $(\nabla\rho,\nabla\Phi)$  and
$L$ is the Laplace operator. See Barcilon et al. (1992)
\cite{Ba1992} and Syganow and von Kitzing (1999) \cite{SyKi1999}
for recent reviews of ion transport in biological media. The same
elliptic-parabolic system had  also surfaced earlier as the
``drift-diffusion'' equations in the semiconductor device design
literature, Selberherr (1984)
\cite{Se1984}.  Similar elliptic-parabolic equations (with
hyper-diffusion, instead of diffusion) also arose
recently as a leading order description of molecular beam
epitaxy \cite{Villain1991}. A variant of the system
(\ref{Keller-Segel})  re-appeared even more   recently as a model
of self-assembly of  particles at nano-scales \cite{MPXB2004}.

\section{Derivation of aggregation equation for\\ finite-size particles
and variable mobility}
In this paper, we modify the class of Debye-H\"uckel equations 
(\ref{Keller-Segel})  with $\epsilon=0$, as a model of the
aggregation of interacting particles of finite size. This problem
is motivated by recent experiments using self-assembly of
nano-particles in the construction of nano-scale devices
\cite{DSA1}-\cite{DSA9}. A colloidal solution of 50nm-size particles is
deposited on a grooved substrate; evaporation and receding contact line
drags the particles into the channels. It has been shown \cite{MPXB2004}
that the most important phenomenon effecting the distribution of particles
in the channels is the capillary interaction between the particles at last
stage of water evaporation, when all water and particles are confined to
the channels.  Examples of particle self-assembly in nano-channels is
shown  in
Fig. 1.  Fundamental principles underlying the mathematics of 
self-assembly at the nano-scales are non-local particle interaction and
nonlinear motion due to variations of mobility at these scales.

 Let us start by deriving  the effective interaction potential between
nano-particles. This will reveal the length scales involved in the problem
and characterize the type of interaction between the particles. 
Due to the effects of surface tension, the air/water interface is deformed
near a particle. On the other hand, water is attracted to the
substrate surface by van der Waals forces. Denote the thickness of the
undisturbed water layer by $H_0$ and let $h$ be the elevation from this
equilibrium level. If we call  the potential of van der Waals
interaction $U_{vdW}(H)$, then for small deviations from the equilibrium
surface the balance of surface tension and van der Waals force gives 
 \begin{equation} 
 \label{vdW}
 \gamma \Delta h=F_0 h,    \hskip1cm 
F_0=-\frac{\partial U_{vdW}}{ \partial H}(H=H_0) 
 \end{equation}  
A typical expression for van der Waals potential includes three terms
\cite{Leshanski-Rubinstein2005}
 \begin{equation} 
 \label{vdWpot0}
 U_{vdW}(H)=\frac{A}{H^3}-\frac{B}{H^\kappa} -C e^{-H/l_0} 
 \end{equation} 
with constants $A$, $B$, $C$, with $\kappa>3$ and $l_0$ being the
interaction length.  A numerical estimate for these values from
\cite{Leshanski-Rubinstein2005} shows that for  $H \sim 50$ nm, the first
term on the right-hand side of (\ref{vdWpot0})  is larger than 
 the second and third terms by factors  $10^9$ and $10^{18}$, respectively.
Here, $A$ is called \emph{Hamaker constant}, which assumes a certain value
for a given pair of materials.%
\footnote{Usually, the 
numerical vallue of $A$ is given in \emph{zJoules}, with \emph{z} being
used for \emph{zepto}$=10^{-21}$. } Then,  (\ref{vdW}) can be re-written
in terms of \emph{capillary length} $l_c$ as 
  \begin{equation} 
 \label{vdW2}
l_c^2 \Delta h- h=0   \hskip1cm l_c=H_0^2\sqrt{\frac{\sigma}{ 3 A}} 
 \end{equation}  
 where we have applied the term \emph{capillary length} even though
the stabilizing potential is due to van der Waals forces and not gravity.
Numerical values for Hamaker constant for water on silicon is about
$30zJ=30 \cdot 10^{-21}J$, surface tension of water/air interface is $0.07
N/m$. Consequently, for $H_0=50$nm, equation (\ref{vdW2})  yields capillary
length $l_c\simeq 300$nm, or several times the particle diameter. 
 
It is now clear that in spite of very different physics and length
scales, the phenomenon of self-attraction of particles at nano-scales
bears not just qualitantive, but also quantitative resemblance to the
well-known \emph{Cheerios effect}: self-attraction of floating bodies due
to surface tension. For floating cheerios, the typical length of
interaction is the  usual capillary length (several mm), which is also of
the order of particle size. It has been shown \cite{Kralchevsky2000, Mahadevan2005} that
two particles separated by a distance $l$ have interaction potential 
proportional  to
$\exp (-l/l_c)$ in one dimension, and $K_0(l/l_c)$ in two dimensions
($K_0(x)$ being the modified Bessel function of the first kind). The force between the particles is then proportional to $K_1(l/l_c)$, and this result holds for both floating and partially submerged particles. Mathematically, these
interaction potentials arise from inversion of Helmholtz operator in
(\ref{vdW2}). 

 These physical considerations show that interaction potentials
proportional to the Green's function for the Helmholtz operator plays a
fundamental role in particle self-assembly across surprisingly many orders
of magnitude. Thus, we shall consistently use this interaction potential in
all our numerical simulations. 
 Nevertheless, we shall keep our framework general
enough that our method of modeling nonlocal interactions among
different particles may be extended to the other physical
problems of interest described in the previous section. 
Our method takes into account the change of mobility due to the
finite size of particles and the nonlocal interaction among the
particles.  The local density (concentration) of particles is
denoted by $\rho$. Suppose the particles interact pairwise via the 
potential
$-G(|\mathbf{r}|)$. The total potential at a point $\bfr$ is
\begin{equation}
\Phi(\bfr)=-\int \rho(\bfr ') G(| \bfr - \bfr') \mbox{d} \bfr'=G * \rho
\label{potential}
\end{equation}
where $*$ denotes convolution. (The minus sign is chosen so
$G>0$ for attracting particles.) The velocity of the
particle is assumed to be proportional to the gradient of the
potential times the mobility of a particle, $\mu$.  The
mobility can be computed  explicitly for a single particle
moving in an infinite fluid. However, when several particles
are present, especially in highly dense states, the
mobility may be hampered by the particle interactions. These
considerations are confirmed, for example, by the observation
that the viscosity  of a dense suspension of hard spheres in
water diverges, when the density of spheres tends to its maximum
value.  Many authors have tried to incorporate the dependence
of mobility on local density by putting $\mu=\mu(\rho)$
\cite{Velazquez2002}. It is common to assume that the mobility
is a function of density, which tends to zero when the density
tends to some maximum value, assumed to be
$\rho_{\hbox{\small max}}=1$, {\em i.e.},
$\mu(\rho_{\hbox{\small max}})=0$. Vanishing mobility
leads to the appearance of weak solutions in the equations,
to singularities and, in general, to massive complications and
difficulties in both theoretical analysis and computational
simulations of the equations. In contrast, the
Keller-Segel model specifies $\mu(\Phi)$, so the mobility is
taken in that case as a function of the concentration of
chemotactic agent (potential) $\Phi$.

The goal of this paper is to suggest and investigate the implications
of the idea that the mobility $\mu$ should depend on the \emph{averaged}
density $\rhobar$, rather than either the potential, or the exact value of
the density at a point. This assumption makes sense from the viewpoints
of both physics and mathematics. From the physical point of view, the
mobility of a finite-size particle must depend on the
configuration of particles in its vicinity. While attempts have
been made to approximate this dependence by using derivatives of
the local density, it is much more natural, in our opinion, to
assume that the local mobility depends on an integral quantity
$\rhobar$ which is computed from the density as $\rhobar=H *\rho$.
Here $H(\bfr)$ is some \emph{filter} function having, in
general, short range compared to the potential $G$. Several
filter functions are possible, with examples being
$H(\bfr)=\delta(\bfr)$ ($\delta$-function),
$H(\bfr)=\exp(-|\bfr| / l)/(2l)$ (exponential, or
inverse-Helmholtz in 1D) or, in $d$ dimensions,  $H(\bfr)=\theta(|\bfr|-l)/(2l)^d$. The
last is the top-hat function, whose value is unity when
$|\bfr|$ is between $-l$ and $l$, and vanishes, elsewhere. The
normalizing factor of $(2 l)^d$ ($d$ being the dimension of space) is introduced so that $ \int_{-
\infty}^{+\infty} H(\bfr)
\mbox{d}\, \bfr =1$. Alternatively, one may assume a filter
function $H(\bfr)$ with zero average:
$ \int_{- \infty}^{+\infty} H(\bfr)\mbox{d}\, \bfr =0$. The
latter assumption may be useful in crystal growth models, where the
mobility does not depend on the absolute level of the
material. Rather, in such models the mobility depends on the
relative positioning of particles. The mathematical analysis
and the reduction property derived in this paper hold,
regardless of the shape of the filter function $H$ and the
potential $G$ is, so long as they are both
\emph{nice} (\emph{e.g.}, piecewise smooth) functions.

\paragraph{Aim of the paper}
The object of this paper is the analysis of the
following continuity equation for density evolution,
\begin{equation}\label{rhoeq}
\underbrace{\
\frac{\partial\rho}{\partial t}
=
-\,
{\rm div}\,\mathbf{J}\
}_{\hbox{Continuity equation}}
\quad\hbox{with}\quad
\underbrace{\
\mathbf{J} =
-\,D\nabla\rhobar -\,\mu(\rhobar)\rho\nabla  \Phi\
}_{\hbox{Particle flux}}
\,.
\end{equation}
Here $D$ is a constant diffusivity, while $\rhobar$ and
$\Phi$ are defined by two convolutions involving,
respectively, the filter function $H$ and the potential $G$,
\begin{equation}
\label{convols}
\rhobar=H*\rho
\quad\hbox{and}\quad
\Phi=G*\rho
\,.
\end{equation}
While this equation preserves the sign of $\rho$, we shall see that it
allows the formation of $\delta$-function singularities even when $D>0$
and in the case of one spatial dimension. 

Alternatively,  system (\ref{rhoeq},\ref{convols}) with $D=0$ can be represented geometrically as
\[
\frac{d}{dt}(\rho\,{\rm dVol})
=0
\quad\hbox{along}\quad
\frac{d\mathbf{x}}{dt}=\mathbf{J}(\rho)
\,,
\]
where the vector field corresponding to the current $\mathbf{J}(\rho)$ is
defined in terms of the pointwise density $\rho\ge0$ and two prescribed
functionals of $\rho$: an energy $E(\rho)$ and a mobility
$\mu(\overline{\rho})\ge0$, as
\[
\mathbf{J}(\rho)=
-\,\rho\mu(\overline{\rho})\nabla
\frac{\delta E}{\delta \rho}
\,.\]
Here $\overline{\rho}=H*\rho\ge0$ is an average density. As a result,
\[
\frac{d}{dt}E(\rho)
=-\int
\frac{1}{\rho\mu(\overline{\rho})}
\,\big|\mathbf{J}(\rho)\big|^2
{\rm dVol}
\le0
\,.
\]
Thus, the flow $d\mathbf{x}/dt=\mathbf{J}$ causes the energy
functional $E(\rho)$ to decrease toward its minimum value and the
flow is Lyapunov stable, provided the energy functional $E(\rho)$ is
sign definite and $\mu(\rhobar)$ is isolated from $0$. We may regard $\mathbf{J}$ as a vector field defining an
infinitesimal action of the diffeomorphisms on the space of positive maps
$\rho$ acting on a manifold $\mathcal{M}$. Remarkably, this action produces
singularities in which the pointwise density concentrates in finite time
on subspaces embedded in the manifold $\mathcal{M}$.

\paragraph{Subcases}
The generality and predictive power of the model
(\ref{rhoeq},\ref{convols}) can be demonstrated by
enumerating a few of its subcases, as follows. When
$H(\bfr)=\delta(\bfr)$ and $G(x)=e^{-|\bfr|/l}$ the system
reduces to the generalized chemotaxis equation
\cite{Velazquez2002}. For the choice $G=\delta'(\bfr)$,
$H=\delta'(\bfr)$ one obtains a modification of the inviscid  Villain model
for MBE evolution \cite{Villain1991} (with extra factor of $\rho$ in the flux).  If the range of $G$
(denoted by $l$) is sufficiently small, we can approximate (at
least formally) the integral operator in $\Phi=G* \rho$ as a
differential operator acting on the density $\rho$, namely,
\[
\Phi=\rho+l \nabla \cdot ( l \nabla \rho)
\,.
\]
As was demonstrated in \cite{MPXB2004}, equation
(\ref{rhoeq}) then becomes a generalization of the viscous
Cahn-Hilliard equation, describing aggregation of domains of
different alloys. In the case that $\mu$ is a constant,
$H(x-y)=\delta(x-y)\,,$ and $G$ is the Poisson kernel,
then equation (\ref{rhoeq}) becomes the drift limit of the
Poisson-Smoluchowski equation for the interaction of
gravitationally attracting particles under Brownian motion
\cite{Ch1939}.

All the particular cases described above require a singular
choice of the functions $G$ and $H$. However, the generalized
functions required in these cases may be approximated with
arbitrary accuracy by using sequences of nice (for example,
piecewise smooth) functions. We shall concentrate on cases
where the functions $H$ and $G$ remain nice, and derive the
results in this more general, more regular, setting. Thus,
regularization in this endeavor introduces additional
generality, which enables analytical progress and makes
numerical solution of the equations easier.

\section{Evolution of density as mass conserving gradient
flow: the nonlocal Darcy's law and energetic considerations}
Let us have another look at our motivation of equations
(\ref{rhoeq},\ref{convols}), this time making a connection with
gradient flows  and thermodynamics. We will show that there is a
naturally defined free energy which remains finite, even when the
solutions cease to be smooth and are replaced by a set of
delta-functions. We show that the
energy remains finite for all choices of
$\mu(\rhobar)$. In contrast, the traditional approach
considering the dependence of mobility on the un-smoothed
density  $\mu(\rho)$ would fail by allowing the energy to
become infinite. This additional energetic argument reinforces
the choice of the regularized  model
in (\ref{rhoeq},\ref{convols}). The calculation will be
performed in arbitrary spatial  dimensions. 

For the energy to be well-defined, we require that the function $G$
describing interaction between two particles is everywhere positive, so
the interaction is always attractive.  In one dimension, we also
require that the kernel function $G$ is symmetric and in
$n$ dimensions that the interaction is central. This assumption is
physically viable for all the historical cases described in
the introduction. Our derivation for the free energy
will remain valid for an \emph{arbitrary} filter function $H$.

\rem{
\begin{equation}\label{cont-eqn}
\frac{\partial\rho}{\partial t}
=
-\,
{\rm div}\,\mathbf{J}
\end{equation}
\rem{Note that the continuity equation preserves the sign of
the mass density $\rho$.
According to Darcy's Law, the velocity is obtained
as the gradient of the pressure,}
The particle flux expression is
\begin{equation}\label{Darcy}
\mathbf{u}=- \nabla{p}
\,.
\end{equation}
We  shall obtain the relation between the pressure
$p$ and density $\rho$ from classical thermodynamics, as
\[
p=\frac{\delta E}{\delta \rho}
\,,
\]
for a free energy functional $E[\rho]$. Let us start with the particular
case when the mobility is constant $\mu(\rhobar) = \mu_0$. Seeking a
situation in which weak
solutions (delta functions) exist, we specify $E[\rho]$ as a \emph{negative}
 norm of the density,
\begin{eqnarray}\label{frerg}
E[\rho]=\frac{\mu_0}{2} \int  \rho  \Phi \, d\,^nx
\quad\hbox{with}
\, \quad \Phi=G*\rho \, \end{eqnarray}
for  some (positive) kernel $G$.  In particular, for
the kernel $G(x)=\exp (-|x|/ \alpha)$, the energy defines  the square of
the norm of the density $\|\rho\|^2_{H^{-1}}$ in the Sobolev
space $H^{-1}$, which remains finite when the solution $\rho$  is either a
smooth function  on a set of delta functions. Hence,  in that particular case, the pressure is given by
\begin{eqnarray}\label{p-def}
p=\frac{\delta E}{\delta \rho}=\mu_0\,\rhobar
\,.
\end{eqnarray}
Thus, when the free energy is a negative Sobolev norm, the
thermodynamic approach singles out the case in which the
pressure is linear in a smoothed mass density. For example,
one could choose the free energy $E[\rho]$ to be the
$H^{-1}$ norm of $\rho$. (We note that delta functions live in
the $H^{-1}$ norm and this motivates us to seek weak solutions
as a sum over delta functions for this situation, in the
example below.) Using several integration by parts, one can
show that this thermodynamic free energy evolves according to
\begin{equation}\label{ddtnorm}
\frac{d}{dt}E[\rho]
=-\int\rho\,|\mathbf{u}|^2
\,d\,^nx\
\,.
\end{equation}
\
\rem{   
\begin{eqnarray*}\label{ddtnorm}
\frac{d}{dt}E[\rho]
&=&
\frac{\mu_0}{2}\frac{d}{dt}\int \rho\rhobar\,d\,^nx
=
-\mu_0\int\rhobar\,\rm{div}\rho\mathbf{u}
\,d\,^nx
\\&=&
\int(\mu_0\nabla\rhobar)\cdot\rho\mathbf{u}
\,d\,^nx
=\int(\nabla{p})\cdot\rho\mathbf{u}
\,d\,^nx
\\&=&
-\int\rho\,|\nabla{p}|^2
\,d\,^nx
=-
\int\rho\,|\mathbf{u}|^2
\,d\,^nx\
\,.
\end{eqnarray*}
} 
So the rate of decay of thermodynamic free energy defines a
(Riemannian) kinetic energy metric in the transport velocity
\cite{Otto2001}.

Let us now turn our attention to
}
We start with equation (\ref{rhoeq}) for mass conservation, in
the case when the mobility $\mu(\rhobar)$ in the particle
flux $\mathbf{J}$ is not constant.  We shall derive a
variant of equation (\ref{rhoeq}) as a gradient flow
in the sense that,
\begin{eqnarray*}
\frac{\partial\rho}{\partial t}
=
-\,
{\rm grad}E\big|_\rho
\,,
\quad\hbox{or}\quad
\Big\langle \frac{\partial\rho}{\partial t}\,,\,\phi\Big\rangle
=
-\,
\Big\langle \frac{\delta E}{\delta\rho}\,,\,\phi\Big\rangle
\,,
\end{eqnarray*}
where $\langle f\,,\,\phi\rangle=\int f\phi\,d^nx$ is the $L^2$
pairing, for a suitable test function $\phi$ and where $E$ is the
following energy,
\begin{equation} \label{frerg}
E[\rho]
=
D\!\!\int\!\! \rho(\log\rho-1)\mbox{d}x
+
\frac{1}{2}\!\!\int\!\!\rho\, \Phi \,\mbox{d}x
\quad\hbox{with}\quad
\Phi=G*\rho
\,.
\label{energy}
\end{equation}
The first term in this expression for the energy $E[\rho]$
decreases monotonically in time for linear diffusion. The
second term defines the $H^{-1}$ norm for the choice
$G(x)=\exp(-|x|/\alpha)$. It may be regarded as a generalization
of this norm for an arbitrary (but positive and symmetric)
function $G$.  Such negative Sobolev norms remain finite,
even when the  solution for the density $\rho$ concentrates into
a set of delta functions.

The variation $\delta E/\delta\rho$ of the free energy
$E[\rho]$ in equation (\ref{frerg}) is
\begin{eqnarray*}\label{Evar1}
\delta E[\rho]
=
\int(D\log\rho + \Phi) \,\delta\rho\,d\,^nx
\,,
\end{eqnarray*}
where we understand $\delta\rho$ as arising from the flow of a
diffeomorphism, as for example in
\cite{HoMaRa1998}. Namely, we specify
\begin{eqnarray*}\label{del-rho}
\delta\rho = -\rm{div}\Big(\rho\mu(\rhobar) \nabla{\Psi}\Big)
\,,
\end{eqnarray*}
for an arbitrary variation of $\rho$ determined by the function
$\Psi$, which is assumed to be smooth.%
\footnote{This specification of the density variation formally
requires the solution to remain differentiable and the product
$\rho\mu(\rhobar)$ not to vanish. In principle, these
conditions would be violated by the formation of weak
solutions, in which the density concentrates into delta-functions.
However, we will verify {\it a posteriori} that the gradient
flow properties of the smooth solutions discussed in this section
are also  preserved for the weak solutions.} 
The corresponding variational derivative is given by, cf.
\cite{Otto2001}
\begin{eqnarray}\label{Evar2}
\delta E[\rho]
&=&
\Big\langle \frac{\delta E}{\delta\rho}\,,\, \Psi\Big\rangle
=
-\int  (D\log\rho +\Phi) \,{\rm div}
\big(\rho \mu(\rhobar)
\nabla{\Psi}\big)d\,^nx
\nonumber\\
&=&-\int \Psi\,
{\rm div} \Big(
\mu (\rhobar)
\big(D\nabla\rho
+
\rho
\nabla\Phi\big)\Big)
\,d\,^nx
\,.
\end{eqnarray}
Consequently, the free energy $E[\rho]$ in equation (\ref{frerg})
produces the following gradient flow:
\begin{eqnarray*}
\Big\langle \frac{\partial\rho}{\partial t}\,,\,\Psi\Big\rangle
=
-
\Big\langle \frac{\delta E}{\delta\rho}\,,\,\Psi\Big\rangle
=
\Big\langle
\rm{div} \Big(
\mu (\rhobar)
\big(D\nabla\rho
+
\rho
\nabla\Phi\big)\Big)
\,,\,\Psi\Big\rangle
\,.
\end{eqnarray*}
The resulting variant of equation (\ref{rhoeq}) is
\begin{equation}\label{rhoeq-var1}
\underbrace{\
\frac{\partial\rho}{\partial t}
=
-\,
{\rm div}\,\mathbf{J}\
}_{\hbox{Continuity eqn}}
\quad\hbox{with}\quad
\underbrace{\
\mathbf{J} =
-\,\mu (\rhobar)
\big(D\nabla\rho
+
\rho
\nabla\Phi\big)\Big)\
}_{\hbox{Modified particle flux}}
\,,
\end{equation}
in which the linear diffusivity of density $\rho$ is modified by
the nonlocal mobility $\mu(\rhobar)$.

\rem{The pressure $p$ is still defined by the variational
derivative $p=\delta E/\delta \rho$.
Since this holds for any
test funstion $\Psi$, we have
\begin{eqnarray}\label{HPeqn}
\frac{\partial\rho}{\partial t}
=
\rm{div}(\rho \mu_(\rhobar) \nabla\Phi)
\,,
\quad\hbox{with}\quad
\rhobar=H*\rho
\,.\end{eqnarray}
}
\paragraph{Energetics}
The free energy $E[\rho]$ in (\ref{frerg})
decreases monotonically in time under the evolution
equation (\ref{rhoeq-var1}). A direct calculation yields,
\begin{equation}\label{erg}
\frac{dE[\rho]}{dt}
=-\int
(D\log\rho +\Phi)\,{\rm div}\,\mathbf{J}
\,d\,^nx
=
-\int
\frac{1}{\rho\mu_(\rhobar)}\,|\mathbf{J}|^2
\,d\,^nx
\,.
\end{equation}
According to this equation, provided $\rho\mu_(\rhobar)>0$,
the rate of decay of free energy $E[\rho]$ given in
(\ref{frerg}) defines a Riemannian  metric in the particle
flux, cf. \cite{Otto2001}.  We shall see that when $D=0$, the
resulting conservative motion of finite-size
particles drifting along the gradient of $\Phi$ leads to a
type of `clumping' of
the density $\rho$ into a set of delta functions. One may check
that this monotonic decrease of energy persists for more general
functions, including weak solutions supported on $\delta$-functions, as
discussed below. 

\rem{  
Many of the same results as those described below will also hold for the
nonvariational equation,
\begin{eqnarray}\label{HPeqn}
\frac{\partial\rho}{\partial t}
=
\rm{div}(\rho \nabla P(\rhobar))
\,,
\quad\hbox{with}\quad
\rhobar=G*\rho
\,,\end{eqnarray}
provided $P\,'(\rhobar)>0$.

} 
\section{Non-vanishing mobility: weak solutions \\ (\emph{clumpons}) and their role in
long-term dynamics in 1D} 
Everywhere in this paper, we shall assume that the potential is purely
attractive, so $G(x)>0$ for all $x$. This assumption is used for two
main reasons. First, it suits the physics of the motivating problem,
namely, mutual attraction of nano-particles \cite{XB2004}. Second,
restricting to purely attractive interactions allows one to separate all
equations  of the type (\ref{rhoeq},\ref{convols}) into two different
classes. The physical property  separating the two classes is whether
the mobility $\mu(\rhobar)$ is strictly isolated from zero, i.e.,  
$\mu(\rhobar) \geq \mu_0 >0$ or one may allow $\mu(\rhobar_0)=0$ for some
$\rhobar=\rhobar_0$. In the  first case, there is nothing to resist the
mutual attraction of the particles and the final state of the system is a
single  $\delta$-function no matter what the form of the mobility
dependence, as long as mobility is strictly isolated from zero. In the
second case, vanishing of the mobility at some maximum density blocks the
motion when the densities become too large and prevents total collapse.
Instead of collapse, this produces isolated \emph{patches} of solutions of
constant density. It is interesting that also in this second case, the
stationary solutions remain \emph{exactly the same} regardless of the
dependence of $\mu$ on $\rhobar$, and we can describe these stationary
solutions  analytically. Moreover, we shall demonstrate that these
solutions are stable and any initial condition separates into a set of
isolated stationary patches of this type.  

This section is devoted to analytical description of dynamics in  the case when $\mu(\rhobar) $ being strictly isolated from zero. 
In simulations, we have assumed $\mu(\rhobar) =1 $ for simplicity, but all our results will remain valid for arbitrary dependence of $\mu$ on $\rhobar$, as long as $\mu$ always remains positive. 

\subsection{Formal weak solution anzatz}
This section considers motion under equations
(\ref{rhoeq},\ref{convols}) in one spatial dimension for the case
of vanishing linear diffusivity, $D=0$. Substituting the
following singular solution ansatz
\begin{eqnarray}\label{Nweak-soln}
\rho(x,t)&=&\sum_{i=1}^Nw_i(t)\delta(x-q_i(t))
\,,\quad
\rhobar(x,t)=\sum_{j=1}^Nw_j(t)H(x-q_j(t))
\,,
\end{eqnarray}
into the one-dimensional version of equation (\ref{rhoeq})
with $D=0$ and integrating the result against a smooth test
function $\phi$ yields
\begin{eqnarray*}\label{weak-soln-eqns0}
&&\int\phi\Big[\rho_t-\Big(\rho\,  \mu(\rhobar) (G*\rho)_x\Big)_x
\Big]dx
=
\int\phi(x)\sum_{i=1}^N
\dot{w}_i\,\delta({x-q_i})dx
\\&&\hspace{2cm}+
\int\phi\,'(x)
\sum_{i=1}^Nw_i\Big(\dot{q}_i+\sum_{j=1}^N w_j(t)\mu \big(\rhobar \big)
G\,'(x-q_j)
\Big)\delta({x-q_i})\,dx
\end{eqnarray*}

Hence, one obtains a closed set of equations for the
parameters $w_i(t)$ and $q_i(t)$, $i=1,2,\dots,N,$ of the
solution ansatz (\ref{Nweak-soln}), in the form
\begin{eqnarray}\label{weak-soln-eqns}
\dot{w}_i(t)=0
\,,\quad
\dot{q}_i(t) = -\sum_{j=1}^N w_j \mu \big(\rhobar \big) G\,'(q_i-q_j)
\end{eqnarray}
where
\begin{equation}
\rhobar= \sum_{m=1}^N w_m  H(q_m(t))
\label{rhobarsingular}
\end{equation}
Thus, the density weights $w_i(t)=w_i(0)=w_i$ are preserved, and the
positions $q_i(t)$ follow the characteristics of the velocity
$\mathbf{u}=-\mu(\rhobar)\nabla G*\rho$ along the Lagrangian
trajectories at $x=q_i(t)$. This result holds in any number of
dimensions, modulo changes to allow singular solutions
supported along moving curves in 2D and moving surfaces in 3D.
Fig.~2, 
demonstrates that the solutions
(\ref{weak-soln-eqns}) do indeed appear spontaneously in a
numerical simulation of equation (\ref{rhoeq}) with $D=0.02$. Hence,
they are ubiquitous and dominate its dynamics. The simulation
started with a smooth (Gaussian) initial condition for density
$\rho$. Almost  immediately, one observes the formation of
several singular \emph{clumpons}, which evolve to collapse
eventually into a single clumpon. Observe that the mass of each
individual clumpon remains almost exactly constant in the
simulations, as required by equation (\ref{weak-soln-eqns}). Also
note that the masses of two individual clumpons add when they
collide and ``clump'' together. Eventually, all the mass becomes
concentrated into   a single clumpon, whose mass (amplitude) is
\emph{exactly} the total mass of the initial condition.

 \subsection{Energy decay close to the final state and estimates for
collapse time}
Normally, systems approaching an equilibrium state tend to evolve
more slowly as they approach the equilibrium. If the rate of
approach diminishes linearly with the distance from equilibrium (in some
sense), for example, one obtains an exponential decay towards the
equilibrium. Alternatively, for finite-time singularities, the rate
tends to diverge to infinity in a  power-law fashion. In contrast to these
familiar examples, the system (\ref{rhoeq},\ref{convols}) approaches the
singularity at a \emph{constant} rate. That is, the rate of approach
to singularity never diverges. Consequently, one may predict the formation 
of singularities and even predict their evolution after they have formed.

This surprising result may be demonstrated by deriving an alternative form
of energy dissipation.  Direct substitution of a single $\delta$-function
for density into (\ref{erg}) is mathematically  ambiguous, as it would
lead to improper operations with $\delta$-functions. Instead, let us
notice that the evolution of the energy $E$ describing the gradient flow
of  (\ref{rhoeq},\ref{convols}) can be  expressed as follows
\[
 \frac{d E}{d t}= -\int G(x-x') \rho(x') \frac{\partial \rho(x)}{\partial
t} \mbox{d} x \mbox{d} x' =  
\] 
\begin{equation}
\begin{array}{cl} 
& +
 \int  \Phi \mbox{div} \left( \rho \mu(\rhobar) \mbox{grad} \Phi \right) 
=-\int \rho \mu(\rhobar) ( \nabla \Phi)^2
\end{array} 
 \label{dEdt}
 \end{equation}
  Note the difference between this formula and (\ref{erg}). Formally,
(\ref{dEdt}) is the same as (\ref{erg}),
  but we can substitute the delta-function ansatz for weak solutions
directly into (\ref{dEdt}).

   Numerical simulations, energetic considerations and physical
intuition suggest that the final state of the system with 
strictly positive mobility $\mu(\rho) \geq \mu_0 >0$ should be a single
clumpon. One may ask how the time necessary to collapse to
this final state varies with the initial conditions. An exact
analytical answer to this question seems out of reach and we shall provide 
a numerical simulation for various initial conditions. However, a 
surprisingly simple and accurate  analytical estimate can be made here.
To start, let us compute the rate of energy dissipated by a
single clumpon, which is the final state of our system. We assume $\rho=M
\delta(x)$, where $M$ is the total mass. Assuming that $H(x)$ is
  regular, $H*\rho (x)=M H(x)$. Let us also assume for simplicity that 
$G(x)=G_0 \exp(-|x|/\alpha)$. Direct computation gives
  \begin{equation}
 \frac{d E}{d t}[\rho=\mbox{clumpon}]= -M \mu \left( M H(0) \right) \left(
\nabla \Phi(0) \right)^2=
 -\frac{M^3}{\alpha^2} \mu \left( M H(0) \right) G_0^2
\label{dEdt0}
\end{equation}
We may now estimate how long it will take for an arbitrary system
to collapse into a single clumpon. If the initial energy of the system is
$E_0$, and initial mass is $M$, the energy of a single clumpon will be
\begin{equation}
E_f=M^2 G_0.
\label{Efinal}
\end{equation}
Then, the approximate  time to collapse to a single clumpon from any
initial conditions is given by combining (\ref{dEdt0}) and (\ref{Efinal}):
\begin{equation}
t_* \simeq \frac{E_0- E_f}{ dE/dt [\rho=\mbox{clumpon}] } =
\frac{E_0-M^2 G_0}{M^3 \mu \left( M H(0) \right) G_0^2 }
\label{collapse}
\end{equation}
In derivation of (\ref{collapse}) we have assumed that $dE/dt$ does not
change much along the trajectories, so the initial conditions are close
enough to a clumpon.  Thus, this estimate for the time of collapse depends
on only two integral quantities: initial energy and mass for
initial conditions close to the final state.

\subsection{Rate of blow up}
The following analysis of the evolution of a density maximum reveals that
the clumping process results from the nonlinear instability of the gradient
flow in equation (\ref{rhoeq}) when $D=0$.  For a particular case
$G=G_0 e^{-|x|/\alpha}$ and $\mu(\rhobar)=1$, one may show that for a high
enough peak,  a density maximum $\rho_m(t)=\rho(x_m(t),t)$ becomes
infinite in finite time.  The motion of the maximum is governed by
\begin{equation} \label{ricatti}
\frac{d}{d t} \rho_m=\frac{1}{\alpha^2}
\left( \rho_m^2-\rho_m \Phi(x_m)  \right) \geq \frac{1}{\alpha^2}
\left( \rho_m^2-\rho_m M \right),
\end{equation}
where $M=\int \rho \mbox{d} x$ is total mass and we have used the fact
that $\Phi$ satisfies
$\Phi-\alpha^2 \Phi_{xx}=\rho$.
The last inequality holds, because $G \leq 1$ is bounded and
$\rho$ is everywhere positive. Thus, if at any point
the maximum of $\rho$ exceeds the (scaled) value of the total mass,
then the value of the density maximum $\rho_m(t)$ must diverge in finite
time. This divergence produces $\delta$-functions in finite time. From
(\ref{ricatti}), the density amplitude must diverge as
$\rho_m \simeq \alpha^2/(t_0-t)$. 
\rem{
To illustrate this divergence,
we have plotted the comparison between the predicted collapse of
$1/\rho_m$ and numerics  in the insert of
Fig.~2. 
}
The formation of
singularities in Fig.~2 occurs both at the maximum, and elsewhere.
The subsidiary peaks eventually collapse with the main peak.

\subsection{Dynamics of inflection points and collapse} 
Let us discuss in more detail the process by which
density singularities are formed. Suppose the the initial condition
contains only one density maximum. Then, one would expect the first
singularity to form at the position of this maximum. As this density
singularity forms, mass flows toward it. This causes a local reduction of
density in the neighboring region from which the mass is flowing. Whenever
this local reduction of density due to the formation of the singularity is
sufficient to form a new maximum in density away from it, then the process
of singularity formation starts again there, and so forth. In principle,
this process could form an infinite number of density singularities.
However, we propose two reasons why the expected number should be finite.
First,  the range of interaction is set by the length scale in $H$. Thus,
one might expect no more clumpons to form than the ratio of the interaction
range to the domain size. Second, in the formation of singularities from
two nearby maxima, one of them may become stronger than the other and
entrain it, thereby suppressing the formation of the second singularity.
In practice, one sees the formation of considerably fewer clumpons than
the number estimated by the ratio of the interaction range to the domain
size. Hence, one may conjecture that the formation of singularities does
involve some competition between neighboring density maxima. This
conjecture could be tested by studying evolution from initial conditions
containing several density maxima separated by varying distances which are
comparable to, or smaller, than the average length scale in $H$.

A quantitative evaluation of this heuristic argument may be obtained by
computing the position  of the inflection point for the averaged density
$\rhobar$. Although no analytical formula for the motion of inflection
point is available, one sees numerically that the inflection point quickly
converges to the position  of the singularity. The evolution of the
inflection point corresponding to the solution from Fig.~2 is shown in
Fig.~3.

\section{Formation of jammed states when mobility $\mu$ approaches zero}
\subsection{Competing length scales of $H$ \& $G$}
An interesting limiting case arises when the scale of non-locality of $H$
is much shorter than the range of the potential $G$.  Formally, this limit
corresponds to $H(x)\rightarrow \delta(x)$. In practice, we  may select a
sequence of piecewise smooth functions $H_\epsilon(x)=\exp
(-|x|/\epsilon)/(2 \epsilon)$ which converge weakly to a $\delta$-function.
For each function $H_\epsilon(x)$ in this sequence, no matter
how small (but positive) the value of $\epsilon$, the exact ODE
reduction (\ref{weak-soln-eqns}) still holds. So, what happens in the
limit of very small epsilon, as $\epsilon\to0$?
\rem{  
As is known \cite{Velazquez2002}, for
$\lim_{\epsilon\to0}H_\epsilon(x)=\delta(x)$, equation (\ref{rhoeq})
exhibits weak solutions which are piecewise constants; either $\rho=0$ or
$\rho=1$.
} 
In investigating this limit, we performed a sequence of numerical
simulations for a fixed choice of the function $G(x)=\exp (-|x|/\alpha)$
while varying $H_\epsilon(x)=\exp (-x/\epsilon)/(2 \epsilon)$ over a
sequence of decreasing values of the ratio $\epsilon/\alpha$.  This
simulation is shown in Fig.~2 for $\epsilon/\alpha=1/10$ and it 
demonstrates the formation of flat clumps of solutions. The mechanism for 
this phenomenon is the following. The vanishing mobility at
$\rhobar=1$ caps the maximal density at $\rhobar=1$ in the
long-term. This leads to the appearance of flat mesas in
$\rhobar(x)$ for large $t$. On the other hand, when $H(x)
\rightarrow \delta(x)$, one finds $\rhobar(x,t) \rightarrow \rho(x,t)$
pointwise in $x$, which forces $\rho(x,t)$ to develop a flat
mesa, or plateau, structure in which the maximum is very close
to unity, as well.  This is precisely what is predicted for the
chemotaxis equation with $H(x)=\delta(x)$ and $\mu(\rho)=1-\rho$
\cite{Velazquez2002}.

In the limit $H \rightarrow \delta$,  model (\ref{rhoeq},\ref{convols})
recovers ordinary diffusion of local density.
This is a singular limit,  because it increases the order of
the differentiation in the equation. Since ordinary
diffusion  is known to prevent collapse in one dimension \cite{Ho2003},
this singular
limit should be of considerable interest for further analysis.

\subsection{Stationary states} 
Numerical simulations of time-dependent problem
(\ref{rhoeq},\ref{convols}) show that the solution
converges to well-defined states  with flat \emph{mesa} peaks when 
$\mu(\rhobar)$ reaches $0$ for some value of $\rhobar$. In the
remainder of this section,  we shall concentrate on the analytical
description of the evolution in this case. For simplicity of formulas, we
shall assume that the critical value of $\rhobar$ is normalized to be
$1$, \emph{i.e}., $\mu(1)=0$. In simulations, we shall take
$\mu=1-\rhobar$. However,  all theoretical results, in particular, exact
expressions for stationary sates, remain true for \emph{arbitrary}
dependence of $\mu(\rhobar)$, as long as $\mu(\rhobar)$ reaches $0$ at
some value of $\rhobar$.  

We distinguish two classes of stationary states, which differ 
mathematically and physically.  Both classes of stationary states will
describe a clump of particles whose density $\rho(x)$ is a generalized
function  with compact support, which we will call a
\emph{patch}. The difference between these classes lies in the
physical reason for the vanishing of local velocity. Remember that 
local velocity is written as $\mathbf{u}=\mu(\rhobar) \nabla \Phi$. The
solution is stationary if $\mathbf{u}$ vanishes at every point
inside the patch. This can be achieved in two ways.  First, $\nabla
\Phi$ may be exactly zero everywhere inside the patch. In this case, the
net force acting on every particle vanishes. These solutions
are called \emph{equilibrium} states. Second, it may also happen that the
mobility $\mu$ vanishes at every point {inside the patch.
Physically, this corresponds to a clump at maximum density whose motion is
prohibited, although the net force on each particle may not be zero. 
These solutions are called \emph{jammed} states. Let us now study both the
equilibrium and jammed  states in more details. As we shall see, an
analytical solution describing each of these states can be found. These
solutions will also elucidate the nature of competing length scales in $G$
and $H$. The stability of these solutions and their selection mechanisms
will also also be demonstrated. 

\subsection{Equilibrium states} 
Let us assume for now that both the potential attraction $G(x)$ and
the filter function $H(x)$ are proportional to the Green's function for
the Helmholtz operator. This is both the case we use in numerical
simulations, and also, incidentally, the only case we have found that
admits analytical solution for stationary states. Thus, we posit
\begin{equation}
\label{GHexp}
G(x)=\exp \left( -\frac{|x|}{\alpha} \right) \hskip2cm H(x)=\frac{1}{2
\beta}\exp \left( -\frac{|x|}{\beta} \right)
\end{equation}
We are looking for stationary states $\rho(x)$ that are weak solutions with
compact support, for which $\rho(x)=0$ if $|x|>L$. While it is impossible
to find a smooth (or even piecewise smooth) function
$\rho(x)$ which is a stationary solution of (\ref{rhoeq}), one 
can find a stationary solution of the following form
\begin{equation}
\rho(x)=w \delta(x+L)+w\delta(x-L)
+
\left\{
\begin{array}{lc}
0, & |x|>L \\
\rho_0(x), & |x| \leq L,
\end{array}
\right.
\label{statsol0}
\end{equation} 
where the value of the  constants $w$ and $L$ are to be determined.

For  (\ref{statsol0}) to be a stationary (weak) solution, the following
conditions must be satisfied:
\begin{equation}
\begin{array}{cc}
\Phi(x)=(G*\rho) (x)=\mbox{const}, & |x|<L    \\
\rhobar(x)=(H*\rho)(x)=1, & x= \pm L.
\end{array}
\label{condstatsol}
\end{equation}
Substitution of (\ref{statsol0}) into the first condition of
(\ref{condstatsol}) gives
\begin{equation}
\label{phicond}
\Phi(x) = w G(x-L)+ w G(x+L) + \int_{-L}^L G(x-y) \rho_0(y) \mbox{d}y
=\mbox{const}
\end{equation}
It is crucial to notice that if $G(x)$ is given by (\ref{GHexp}), then the
integral in (\ref{phicond}) must yield a function proportional to a sum of
exponentials, which is only possible if $\rho_0(x)$ is a constant:
$\rho_0(x)=\rho_0$. Performing the integral gives
 \begin{equation}
\label{phicond2}
\Phi(x)=w e^{-|x-L|/\alpha}+ w e^{-|x+L|/\alpha} + \rho_0 \alpha \left(
2-e^{-|x-L|/\alpha}- e^{-|x+L|/\alpha}
\right) =\mbox{const}
\end{equation}
which requires
\begin{equation}
\label{wcond}
w=\rho_0 \alpha.
\end{equation}
Enforcing the second condition in (\ref{condstatsol}) will
determine $\rho_0$ as a function of $L$. As in (\ref{phicond}), we find an
analytical expression for $\rhobar$
\begin{equation}
\rhobar(x)=\frac{w}{2 \beta} \left(  e^{-|x-L|/\beta}+
e^{-|x+L|/\beta}\right)  + \frac{\rho_0}{2}
 \left( 2-e^{-|x-L|/\beta}- e^{-|x+L|/\beta} \right)
\label{rhobarcond}
\end{equation}
Since (\ref{rhobarcond}) assumes identical values at $x=\pm L$, we need
only check that $\rhobar(x=L)=1$.
Substitution of $x=L$ into (\ref{rhobarcond}) gives
\begin{equation}
\rho_0 \left[  \alpha \left( 1+\exp\left( - 2 L/\beta \right) \right) +
\beta \left( 1-\exp\left( - 2 L/\beta \right) \right) \right]=2 \beta
\label{rho0cond}
\end{equation}
In principle, equation (\ref{rho0cond}) already determines the density
inside the patch as a function
of length. It is, however, more practical to determine the length as a
function of total mass of the clumpon
$M$. The mass contained in the solution (\ref{statsol0}) is
$$M=\rho_0 L+2 w =\rho_0 (L + 2 \alpha).$$
Thus, we find an implicit  condition for length $L$ as a function of patch
mass $M$:
\begin{equation}
M \left[  \alpha \left( 1+\exp\left( - 2 L/\beta \right) \right) +
\beta \left( 1-\exp\left( - 2 L/\beta \right) \right) \right]=2 \beta (L+2
\alpha)
\label{lcond}
\end{equation}
These solutions are shown on Fig.~5. 
It is interesting to note that (\ref{lcond}) defines a non-negative length
$L$ only if the mass $M$ is sufficiently large. This is perfectly
physical: if the mass is small, there is no reason for the solution to
``spread out", so a single $\delta$-function is produced. Incidentally,
this will happen if the mass of the clumpon $M=\int \rho \mbox{d} x$ is
such that $max_x \rhobar<1$, i.e., $M<2 \beta$. 
Physically, we expect that equilibrium solutions are unstable. Indeed,
imagine a set of particles on a line  positioned at an equal distance from
each other. If the force between each pair of particles is repulsive,
this  configuration is stable, and the real part of all eigenvalues of the
linearized problem is negative. However, if the force between the
particles changes to purely attractive without changing the particle
positions, this leads to the sign change of the eigenvalues
which correspond to a stable situation. Thus, these equilibrium states
with fixed gaps between the particles ``held together" by purely
attractive forces must be  unstable, and must exhibit large values for
the real parts of the unstable eigenvalues. Numerical analysis of linear
stability  of equilibrium states  in the continuum description confirms
this intuitive physical picture. Namely, there exists a set of eigenvalues
with large positive real parts. In addition, a fully nonlinear
time-dependent simulation of (\ref{rhoeq},\ref{convols}) starting with
initial conditions corresponding to even slightly perturbed equilibrium
states shows very rapid deviation away from equilibrium. Therefore,
equilibrium states are \emph{unstable} and will never be realized in
nature.  

\subsection{Jammed states} 
The derivation of the jammed states is rather similar to that
for the equilibrium states, so only a brief discussion of it will be
provided.  Let us again assume that stationary states $\rho(x)$ are weak
solutions with compact support, and
$\rho(x)=0$ if $|x|>L$: 
\begin{equation}
\rho(x)=w \delta(x+L)+w\delta(x-L)
+
\left\{
\begin{array}{lc}
0, & |x|>L \\
\rho_0(x), & |x| \leq L
\end{array}
\right.
\label{statsoljammed0}
\end{equation}
Now, for solution (\ref{statsoljammed0}) to be a jammed solution, we need
just one condition: 
\begin{equation}
\rhobar(x)=(H*\rho)(x)=1, \hskip2cm |x| \leq \pm L.
\label{condstatsoljammed}
\end{equation}
Substituting (\ref{statsoljammed0}) into condition 
(\ref{condstatsoljammed}) gives
\begin{equation}
\label{rhobarcondjammed}
\rhobar(x) = w H(x-L)+ w H(x+L) + \int_{-L}^L H(x-y) \rho_0(y) \mbox{d}y
=1
\end{equation}
Again, it is essential that $H(x)$ is given by (\ref{GHexp}), so the
integral in (\ref{phicond}) must be proportional to a sum of
exponentials, which is only possible if $\rho_0(x)$ is a constant:
$\rho_0(x)=\rho_0$.  Equation (\ref{rhobarcondjammed})  transforms
to
 \begin{equation}
\label{rhobarcond2}
\rhobar(x)=\frac{w}{2 \beta} 
\left( e^{-|x-L|/\beta}+ e^{-|x+L|/\beta}\right) 
+ \frac{\rho_0}{2} \left(
2-e^{-|x-L|/\beta}- e^{-|x+L|/\beta}
\right) =1
\end{equation}
which requires
\begin{equation}
\label{wcondjammed}
w=\beta    \hskip2cm   \rho_0=1
\end{equation}
The mass contained in the solution (\ref{statsol0}) is
$$M=\rho_0 L+2 w =L + 2 \beta.$$
The jammed states are illustrated on Fig.~6. 

To understand the nonlinear stability of jammed states, let us again
appeal to the physical picture of particles on a line. \emph{Jammed}
states correspond to a set of finite-size particles pressed  tightly
together. Physical intuition tells us that such a state should be
favorable for purely attractive forces between pairs of particles.
Numerical analysis of linear stability of \emph{jammed} states in the
continuum approximation confirms this: the real part of the spectrum is
isolated from $0$ by $-D$, where $\lambda=-D$ is the limiting  point of
spectral sequence. Thus, $D$ should determine the time scale for
convergence to stationary solutions. That is, the rate of convergence to
stationary solution is given by the time scale $\tau \sim 1/D$. 

To verify these predictions, a fully nonlinear simulation starting with a
smooth initial conditions in the form of Gaussian peak has been performed.
The results of this simulation are given in Fig.~7. For $D=0.01$ used in
the simulation, the solution becomes  practically stationary after $t \sim
100$. The small ``bumps" at the position of the $\delta$-functions in
density correspond to a mismatch in the stationary solutions,  which were
derived for $D=0$. The amplitudes of the ``bumps" tend to zero when $D$
becomes very small.

\section{Jammed stationary states in two dimensions} 
\subsection{Exact jammed states in two dimensions \label{sec:2dsols}}

While this paper focuses primarily on the evolution of particles in
one dimension, the real-world technological importance of self-assembly
processes requires us to derive and analyze stationary solutions of our
model in two dimensions.  Numerical simulations of particle dynamics
\cite{Japanese} show that there is a tendency for the formation  of
isolated clumps of either roughly circular shape, or, for high particle
densities, roughly circular or elliptical voids in fully dense areas. To
model these results, let us assume that the mobility $\mu(\rhobar)$
vanishes at critical density $\rhobar=1$, and $\rho$ and $\rhobar$ are 
connected through the two-dimensional Helmholtz operator
\begin{equation} 
\label{2DHelmholtz}
\rhobar-\beta^2 \Delta \rhobar=\rho
\,,\quad\hbox{so that}\quad 
\rhobar=H*\rho
\,.
\end{equation}
Inspired by these results we use the intuition developed in one
dimension to seek jammed solutions in two dimensions which are fully
dense $\rhobar=1$ inside an area $D$  with additional distributed
$\delta$-function for density on the boundary  $\partial D$, and $\rho=0$
in the exterior of $D$.  Remarkably, when $H$ is the Green's function for
the Helmholtz operator, an exact analytical expression for several
possible shapes of $D$ can be found. Again, we shall emphasize that our
results will hold for an \emph{arbitrary}  functional dependence
$\mu(\rhobar)$ as long as the mobility vanishes for some value of
$\rhobar=\rhobar_*$. For  convenience, we have again rescaled the
critical value to be $\rhobar_*=1$.

Consider an orthogonal coordinate system $(\xi,\eta)$ for
which the solution of Helmholtz equation separates variables.
Suppose, in addition, that the boundary of $D$ corresponds to one of the
coordinate lines $\xi=\xi_0$, and interior of $D$ is given by $\xi<\xi_0$
Then, we may seek the solution in the exterior of $D$ in the form
$\rhobar(\xi,\eta)=F(\xi/ \beta) G(\eta / \beta)$. Boundary conditions for
the exterior of $D$ are $\rhobar \rightarrow 0$ as 
$\xi \rightarrow +\infty$, and boundary conditions at $\xi=\xi_0$ are
obtained by continuity of $\rhobar$ at the boundary. Since for jammed
states $\rhobar=1$ in the interior, the continuity condition gives
$\rhobar=1$ on the boundary. Therefore, separable solutions must
take the form 
$\rhobar(\xi,\eta)= F(\xi/\beta)/F(\xi_0/\beta)$. Since the system of
coordinates is orthogonal and $\rho=1$ in the interior of $D$ from
(\ref{2DHelmholtz}), we can compute the amplitude of $\delta$-function for
density $\rho$ at the boundary $\xi=\xi_0$, which comes out to be $-\beta 
F'(\xi_0/\beta)/F(\xi_0/ \beta)$. Thus, the exact solution for the case
when separation of variables of Helmholtz equation is possible and the
boundary is given by $\partial D=\left\{ \xi=\xi_0 \right\}$ is
\begin{equation} 
\label{2Dsolrho} 
\rho(\xi,\eta)=\left\{ 
\begin{array} {lc} 
1, & (\xi,\eta) \in \mbox{int}D \\
-\delta(\xi-\xi_0)\beta  F'(\xi_0/\beta) /F(\xi_0/\beta) 
& (\xi,\eta) \in \partial D \\
0, & (\xi,\eta) \in \mbox{ext}D 
\end{array} 
\right. 
\end{equation} 

\begin{equation} 
\label{2Dsolrhobar} 
\rhobar(\xi,\eta)=\left\{ 
\begin{array} {lc} 
1, & (\xi,\eta) \in \mbox{int}D \\
F(\xi/\beta)/F(\xi_0/\beta), & (\xi,\eta) \in \mbox{ext}D 
\end{array} 
\right. 
\end{equation} 
To be specific, in Table~1 we enumerate all cases for which separation of
variables in Laplace/Helmholtz  equation is possible. As it is known,
Helmholtz equation in two dimensions is separable in four coordinates
only: Cartesian (which we shall not consider here), cylindrical, elliptic
cylindrical and parabolic cylindrical \cite{MorseFeshbach}. Each of these
cases selects a particular shape of the patch, as well as a particular form
of the function $F(\xi)$. 
\begin{table}[h]
\begin{tabular} {|c|c|c|}
\hline 
\emph{Coordinates} & \emph{Shape of $D$} & 
\emph{ Functional form of $F(\xi)$}  \\
\hline 
Cylindrical &  Circle &  Bessel function $K_0(r)$  \\
\hline 
Elliptic cylindrical & Ellipse & Modified Matthew Function \\
\hline 
Elliptic cylindrical & Hyperbolae &  Matthew Function \\
\hline 
Parabolic cylindrical & Parabola & Parabolic Cylinder Function \\
\hline
\end{tabular}
\caption{A summary of exact jammed solutions and 
corresponding shapes in two dimensions } 
\end{table} 

Of course, similar results may be obtained in three dimensions, where
eleven coordinate systems exist for which the Helmholtz differential
equation is separable. We shall not go into details here as, first of all,
the generalization is straightforward, and, second, we are only interested
here in one- and two-dimensional self-assembly. Three-dimensional analogues
of self-assembly  may be important for some models arising in mathematical
biology which are discussed at the end of Sec.\ref{sec:conclusions}.  }
\subsection{Irregular shapes: asymptotic result} 

For arbitrary shape of domain $D$, it is no longer possible to find an
exact expression for densities  $\rho$ and $\rhobar$. However, we may find
an \emph{asymptotic} result which shows that $\delta$-function density on
the boundary must be a slowly-varying function of the boundary coordinates
in  the case when $\beta$ -- the range of $H$ -- is small compared to the
size of the patch. We consider a patch of an arbitrary  simple-connected
shape $\Omega$ with a smooth boundary and introduce the coordinate $\eta$
along the boundary of the patch $\partial \Omega$. Considerations of  one
dimensional case indicate that the patch is stable if it is jammed, i.e.
$\mu=0$ everywhere inside the patch, which leads to equation $\rhobar=1$
inside the patch. For now, we consider $H(x)$ to be inverse Hemholtz
operator, thus $\rhobar$ satisfies (\ref{2DHelmholtz}). Thus, inside the
patch, we necessarily have $\rho=1$. The decay of $\rhobar$ close to the
boundary should be compensated by the presence of delta-function
concentration at the boundary. Let us assume that the boundary is
smooth. If $\xi$ is the coordinate locally perpendicular to the boundary
and the boundary is at $\xi=\xi_0$, then we assume that the density near
the boundary  has the form:
 \begin{equation} 
 \rho(\xi \sim \xi_0)= f(\eta) \delta(\xi-\xi_0)+\mbox{ind}_\Omega(\mathbf{r}),
 \label{deltadistribution} 
 \end{equation} 
 where $\mbox{ind}_\Omega(\mathbf{r})$ is the indicator function which 
equals unity. if $\mathbf{r} \in \Omega$ and vanishes otherwise.  From the
condition $\rhobar=1$ and $\rho=1$ in the interior of $\Omega$ we obtain
an integral equation for the strength $f(s)$: 
\begin{equation} 
\int _{\partial \Omega} f(s) H(\mathbf{r}-\mathbf{r}(s') )  \mbox{d} s'+ 
\int_{\Omega} H(\mathbf{r}-\mathbf{r'})  \mbox{d} \mathbf{r}'=1
\label{intf} 
\end{equation} 
If $\mathbf{r}$ is farther than $\beta$ away from the boundary, then
only the contribution from the second integral is relevant. However, due
to the extremely short range of $H$, this integral is equal to unity and 
equation (\ref{intf}) gives an identity. Thus, we only need to investigate
equation (\ref{intf}) in the case when  the distance between $\mathbf{r}$
and the boundary is of order $\beta$ or less. Suppose this distance  is $d
\beta$ with $d>0$ being a constant of order unity or less. Since
$H(\mathbf{r}'-\mathbf{r})$ decays rapidly away from $\mathbf{r}$ for
distances larger than $\beta$, we can approximate the slowly varying
function $f(s')$ by its value at $s$, and the integral (\ref{intf}) as 
\begin{equation} 
f(s) \int  _{\partial \Omega} H(\mathbf{r}-\mathbf{r}(s') )  \mbox{d} s'+ 
\int_{\Omega} H(\mathbf{r}-\mathbf{r'})  \mbox{d} \mathbf{r}'=1
\label{intf2} 
\end{equation} 
Since we are only interested in the immediate neighborhood of the point
$\mathbf{r}(\eta)$, we will now  assume that the boundary is locally
straight and vertical, and the interior of the patch is to the right of
the boundary. We introduce the local coordinates $x=\beta \xi$ and
$y=\beta \eta$ centered at the point $\mathbf{r}$, so the boundary is at
$x=d$ and the $x-$axis is pointing towards the boundary.  Note the exact
form of the kernel $H$ 
\begin{equation} 
H(\mathbf{r} ) =\frac{1}{2 \pi \beta^2} K_0\left( \frac{|\mathbf{r}|}{\beta}  \right).
\label{K0} 
\end{equation} 
We split up the second integral in (\ref{intf2})  and perform
the integration exactly by going to polar coordinates (and skipping
algebraic details): 
\begin{equation} 
\int_{\Omega} H(\mathbf{r}-\mathbf{r'})  \mbox{d} \mathbf{r}' \simeq
\frac{1}{2 \pi}  \int_{x<d} K_0 \left( \sqrt{x^2+y^2} \right) \mbox{d}x \mbox{d}y = 
\label{splitint}
\end{equation} 
\[
 =1-\frac{1}{2 \pi}  \int_{x>d} K_0 \left( \sqrt{x^2+y^2} \right) \mbox{d}x \mbox{d}y =1-\frac{1}{2} e^{-d}\]
\rem{  
Going to polar coordinates, we can perform the last integral exactly: 
\[ 
\int_{x>d} K_0 \left( \sqrt{x^2+y^2} \right) \mbox{d}x \mbox{d}y =\int_{r>d} \mbox{d} r \int_{-\phi(r)}^{\phi(r)} K_0(r) r \mbox{d} \phi=
2\int_{r>d} K_0(r) r \phi(r) \mbox{d} r, 
\] 
where $\phi(r)=\mbox{arccos}(d/r)$. This integral can be performed
analytically to give 
\[ 
2\int_{r=d}^{\infty}K_0(r) r \mbox{arccos}\left(\frac{d}{r} \right)  
\mbox{d} r=\pi e^{-d},
\] 
\[ 
-\frac{1}{2 \pi}  \int_{x>d} K_0 \left( \sqrt{x^2+y^2} \right) 
\mbox{d}x \mbox{d}y =\frac{1}{2} e^{-d} 
\] 
so 
\begin{equation} 
 \int_{x>d} H(x,y) \mbox{d}x \mbox{d}y=\pi e^{-d} 
 \label{I2} 
 \end{equation} 
} 
\noindent 
 Let us now compute the first integral in (\ref{intf2}). We have: 
 \[ 
 \int H(\mathbf{r}(s)-\mathbf{r}(s') )  \mbox{d} s' = \frac{1}{2 \pi \beta}  \int_{y=-\infty}^{+\infty}
  K_0\left(\sqrt{d^2+y^2} \right) \mbox{d} y= 
  \frac{1}{2 \beta} e^{-d} 
  \] 
  \rem{ 
  The last integral can also be computed analytically and we get 
  \[ 
   \int H(\mathbf{r}(s)-\mathbf{r}(s') )  \mbox{d} s=\beta \pi e^{-d} 
   \] 
   } 
   Thus, equation (\ref{intf2}) becomes 
   \begin{equation} 
   \label{intf3} 
   f(\eta) \frac{1}{2 \beta} e^{-|\mathbf{r-r'}|/\beta}
   + 1-\frac{1}{2} e^{-|\mathbf{r-r'}|/\beta} =1 
   \end{equation} 
so the answer is simply 
\begin{equation} 
f(\eta)=\beta.
\label{fanswer} 
\end{equation} 
Thus, for a smooth boundary, jammed states are obtained by the same
principle as in one dimension: the  densely packed state $\rho=\rhobar=1$
inside the domain is surrounded by a $\delta$-function layer of strength
$\beta$.  At first glance, the asymptotic answer (\ref{fanswer}) seems
not correspond to the exact answer (\ref{2Dsolrho}) for circular, elliptic
and parabolic shapes. Consider, however, the case of a circular patch as an
example.  Assume that the boundary of the patch being at $r=r_0$.
According to (\ref{2Dsolrho}), the strength of $\delta$-function on the
boundary is 
\[ 
f=-\beta \frac{K_0' \left( r_0/\beta \right )}{K_0 \left( r_0/\beta \right)} = 
\beta \frac{K_1 \left( r_0/\beta \right )}{K_0 \left( r_0/\beta \right)} 
\] 
The ratio of Bessel functions converges rapidly to  $1$ for $\beta \rightarrow 0$. In fact, this 
ratio is extremely close to $1$ already when $\beta < r_0 $. 
Since $\beta$ is assumed  small, for a patch of any reasonable
size (larger than several units of $\beta$), the approximate result
(\ref{fanswer}) is valid to high accuracy (exponential in $\beta$).  In
general, we can only postulate that the accuracy of (\ref{fanswer}) 
should be $O(\beta^2)$, since we neglected the local curvature of the
surface. Introduction of local curvature effects will change the
asymptotic result (\ref{fanswer}), but this correction is bound to be
small  (of the order $\beta^2$) when $\beta \rightarrow 0$.  
\subsection{Numerical simulation of states in two dimensions} 

To illustrate these exact and asymptotic results, we have performed fully
nonlinear numerical simulations for $\alpha=1$, $\beta=0.1$ and $D=0.01$
in two dimensions. The results of this simulations are shown in Fig.~8.
gaussian radial distribution. Evolution due to (\ref{rhoeq},\ref{convols})
deforms this shape  into a flat-top elliptical shape, which is reminiscent
of  the solutions derived in Sec.\ref{sec:2dsols}. We  conjecture that
elliptical shapes are more stable than circular shapes, although a 
detailed analysis of the two-dimensional stability and selection mechanism
has yet to be completed.  
\section{Conclusion, open problems and further applications \label{sec:conclusions}}
A new model was proposed and analyzed for the
collective aggregation of finite-size particles driven by the force of
mutual attraction.
Starting from smooth initial conditons, the solution
for the particle density in this model was found to collapse into a set of
delta-functions (clumps), and the evolution equations for the dynamics of
these clumps were computed analytically. The energy derived for this model
is well defined even when density is  supported on
$\delta$-functions. The mechanism for the formation of these
$\delta$-function clumps is the nonlinear instability governed by the
Ricatti equation (\ref{ricatti}), which causes the magnitude of any density
maximum to grow without bound in finite time.
\rem{ 
In this letter, we proposed and analyzed a new model for the
collective aggregation of finite-size particles driven by the force
of mutual attraction. The model exhibits the formation
of singularities.
} 
At first sight, it may seem that the emergence of
$\delta$-function peaks in the solution might be undesirable and
perhaps should be avoided. However, these $\delta$-functions
may be understood as clumps of matter, and the model guarantees that
any solution eventually ends up as a set of these clumps.
Subsequently, further collective motion of these clumps may be
predicted using a (rather small-dimensional) system of ODEs,
rather than dealing with the full non-local PDEs.
The question of how many clumps arise from a given initial condition
remains to be considered. One may conjecture that clump formation is
extensive; so that each clump forms from the material within the range of
the potential $\Phi$, determined by $G$. On a longer time scale, the clumps
themselves continue to aggregate, as determined by the collective
dynamics (\ref{weak-soln-eqns}) of weak solutions (\ref{Nweak-soln}). This
clump dynamics is also a gradient flow; so that eventually only one clump
remains.

A comparison with point vortex solutions of Euler's equations for
ideal hydrodynamics in two-dimensions may be made here. Prediction
of the incompressible motion of an ideal fluid is governed by a
set of nonlinear PDE in which pressure introduces non-locality.
A drastic  simplification of motion occurs, when all the
vorticity is concentrated in delta-functions (point vortices)
\cite{Saffman-book}. The motion of point
vortices lies on a singular invariant manifold: if
started with a set of point vortices, the fluid structure will
remain a set of point vortices. However, a smooth initial
condition for vorticity \emph{does not} split into point
vortices under the Euler motion. In the present model, though,
the physical attraction drives any initial distribution of density towards a
set of delta-functions, so one is \emph{guaranteed} to obtain
effectively finite-dimensional behavior in the system after a
rather short initial time. Of course, this time depends of the precise form
of the long range attraction.

Because two scales are present in the smoothing functions $H$ and $G$, the
effects of boundary conditions warrant further study. For example,
 the $H \rightarrow \delta$ limit should also allow
formation of boundary layers. These boundary issues were avoided
in the present treatment
by using periodic boundary conditions.  However, it would be natural in
some physical situations to apply, e.g., Neumann boundary conditions
to the particle flux $\mathbf{J}$.  The issue of boundary conditions will
arise and must be addressed on an individual basis in specific applications
of these equations in physics, chemistry, technology and biosciences.

Our approach involving a variational principle, energy and competition of
length scales is relevant for many areas of science. In particular, we
have recently learned that our variational approach is applicable to
bio-sciences, in particular, to the theory of insect swarming. In a recent
work, Topaz {\em et al.} \cite{TBL2005} have discovered the formation of
isolated patches of matter (\emph{swarms}) in  a variant of 
(\ref{rhoeq},\ref{convols}) with a \emph{nonlinear} instead of
\emph{nonlocal} diffusion. The variational energy (\ref{frerg}) for
their model was 
\[ 
E[\rho]=\int \rho G* \rho-\frac{D}{2} \rho^2 
\] 
This case could be considered as an extension of our model for $H(x)=
\delta(x)$, $\mu(\rhobar)=1$  and 
\[ 
E[\rho]=\int \rho G* \rho-D Q(\rhobar)  
\] 
with $Q'(\rhobar) \geq 0$, in particular, $Q(\rhobar)=\rhobar^2 /2$.
Alternatively, a variant of this nonlinear  diffusion model can be derived
if we start with $\mu(\rhobar)=1-\rhobar$ and $D=0$.  We can then rewrite
(\ref{rhoeq}) as  
\[ 
\frac{\partial \rho}{\partial t}+\mbox{div} \left( \rho \nabla \Phi \right)= \mbox{div} \left(  \rho \rhobar \nabla \Phi \right),  
\] 
which, again,  is a modification of swarming models considered in \cite{TBL2005} with non-local diffusion. We emphasize that, in our opinion, nonlocality in diffusion is advantageous, since it allows for simple generalized solutions in terms of constants and $\delta$-functions. These generalized solutions dominate both the dynamics and statics of the problem and greatly simplify the analytical treatment.

{\bf Acknowledgments.}  We thank S.~R.~J.~Brueck, P. Constantin, B. J. Geurts, J.~Krug and
E. S. Titi for
encouraging discussions and correspondence about this work. DDH is 
grateful for partial support by US DOE, under contract W-7405-ENG-36 for
Los Alamos National Laboratory, and Office of Science ASCAR/AMS/MICS. VP was partially supported by funding from US DOE, DE-FG02-04ER-46119 and Petroleum Research Grant 40218-AC9.

\newpage
\begin{center} 
{\bf \large Figure Captions} 
\end{center} 
\begin{enumerate} 
\item[Figure1.]
Scanning Electron Microscope (SEM) figures of self-assembly of particles in nano-channels, courtesy of S.~R.~J.~Brueck and D.~Xia. 
Note fully dense clumps separated by voids, 
evident in cases B, C and F.  Cross-width of the channels as well as distance between the channels is 100 nm.
\item[Figure2.] 
  Numerical simulation demonstrates the 
emergence of particle clumps, showing formation of density
peaks in a simulation of the initial value problem
for equation (\ref{rhoeq}) using smooth initial conditions for
density with $l=1$, $G(x)=H(x)=\exp (-|x|) $,
$\mu(\rhobar)=1$. The vertical coordinate represents 
$\rhobar=H*\rho$, which remains finite even when the density forms $\delta$-functions. 
 \item[Figure 3.] 
Position of inflection point for simulation shown on Fig.~2. The dashed line in the bottom shows the minimum value allowed by the finite resolution of the mesh. In this case particular case, this minimum value is equal  0.1 with  the length of the 
interval being 10.
\item[Figure 4.] Evolution of a Gaussian initial
condition for $\rho$(x,0) with $\mu(\rhobar)=1-\rhobar$ and
$H=\exp(-|x|/\epsilon)/(2 \epsilon)$ where $\epsilon=\alpha/10$. The
solution quickly forms a plateau of maximal possible density
($\rho_{\hbox{\small max}}=1$).  
\item[Figure 5.] 
Stationary equilibrium solution for $\alpha=1$, $\beta=0.1$ and $L=1$. Top
$\rho(x)$ (with representation of $\delta$-functions at
$x=\pm L$ as vertical red lines).  Middle: $\Phi=(G*\rho)(x)$ for the same solution.
Bottom: $\rhobar=H*\rho$ for the same solution. 
\item[Figure 6.] 
Stationary jammed states for $\alpha=1$, $\beta=0.1$ and $L=1$. Top
$\rho(x)$ (again,  $\delta$-functions at
$x=\pm L$ are represented by vertical red lines).  Middle: the potential $\Phi=(G*\rho)(x)$.
Bottom: $\rhobar=H*\rho$. 
\item[Figure 7.] 
Convergence to analytic equilibrium solution (circles)  for different times 
(colored solid lines, see legend). In simulation, $D=0.01$, $\alpha=1$, $\beta=0.1$ 
\item[Figure 8.] 
Two-dimensional evolution of a gaussian initial profile for consecutive
times with $D=0.01$, $\alpha=1$, $\beta=0.1$ 
\end{enumerate}

\end{document}